\documentclass{aa}  
\usepackage{natbib} 
\usepackage{graphicx}
\usepackage{multirow}
\usepackage{txfonts}
\usepackage{xcolor}
\usepackage{ulem}
\newcommand{\cc}[1]{\textcolor{black}{#1}}
\newcommand{\as}[1]{\textcolor{black}{#1}}

\usepackage[colorlinks,allcolors=blue]{hyperref}
\begin{document} 
\title{Tidal interactions shape period ratios in planetary systems with three-body resonant chains}
\titlerunning{Tidal interactions shape period ratios in resonant chains}

\author{C. Charalambous, J. Teyssandier
  \and
  A.-S. Libert}

\institute{naXys, Department of Mathematics, University of Namur, Rue de Bruxelles 61, 5000 Namur, Belgium\\
\email{carolina.charalambous@unamur.be}
   }

\date{Received XXX; accepted YYY}

\abstract
{}
{These last years several Systems with Tightly packed Inner Planets (STIPs) in the super-Earth mass regime have been discovered harboring chains of resonances. It is generally believed that planet pairs get trapped in mean-motion resonance (MMR) during the migration phase in the protoplanetary disk, while the tides raised by the host star provide a source of dissipation on very long timescales. In this work, we aim to study the departure from exact commensurabilities observed among the STIPs which harbor 3-planet resonances and analyze how tides play an important role in shaping the resonance offsets for the STIPs.}
{We analyzed the resonance offsets between adjacent pairs for five multi-planetary systems, namely Kepler-80, Kepler-223, K2-138, TOI-178, and TRAPPIST-1, highlighting the existence of different trends in the offsets. On the one hand, we derived analytical estimates for the offsets, which confirm that the departure of the planetary pairs from the nominal MMRs are due to the 3-planet resonant dynamics. On the other hand, we performed N-body simulations including both
orbital migration and tidal dissipation from the host star with simple prescriptions in order to test the effectiveness of this mechanism at shaping the observed trend in the offsets, focusing our study on the preservation of the resonant patterns in the different systems with the same general setup.}
{We found that the trends in the offsets of the five detected systems can be produced by tidal damping effects, regardless of the considered value for the tidal factor. It is a robust mechanism that relaxes the system towards equilibrium while efficiently moving it along 3-planet resonances, which induces the observed resonance offset for each planet pair. In addition, we showed that for Kepler-80, K2-138, and TOI-178, the amplitudes of the resonant offsets can also be reproduced with appropriate tidal factor, for the estimated age of the systems.} 
{}


\maketitle
%
\section{Introduction}

Since the discovery of the hot Jupiters, it is widely accepted that planets do not form in their present location, but farther from the host star, and subsequently migrate inwards while the protoplanetary disk is still present. During the (convergent) migration process, adjacent planets are expected to get trapped in mean-motion resonance (MMR)\footnote{Two planets are said to be in a $(p+q)/p$ MMR, with $p,q \in \mathbb{N}$, if the mean motions $n_1$ and $n_2$ of the planets satisfy $(p+q)n_2 - p n_1\sim 0$ (i.e., commensurability of the orbital periods $P_1$ and $P_2$) and at least one of the associated resonant angles combining the mean longitudes and the longitudes of the pericenters librate around a fixed value.}, and when three or more planets are linked by MMRs, a resonant chain can be formed. Resonant chains are of particular importance for the STIPs \citep[Systems with Tightly packed Inner Planets, ][]{2012DPS....4420004R} due to the close proximity between the planets. Up to date, six exoplanetary systems host planets which are confirmed in resonant chains, namely GJ-876 \citep{2010ApJ...719..890R}, Kepler-223 \citep{2016Natur.533..509M}, Kepler-80 \citep{2014ApJ...784...44L,2014ApJ...784...45R}, K2-138 \citep{2018AJ....155...57C}, TOI-178 \citep{2019A&A...624A..46L}, and TRAPPIST-1 \citep{2016Natur.533..221G,2017Natur.542..456G}, while HR 8799 \citep{2008Sci...322.1348M} and Kepler-60 \citep{2013MNRAS.428.1077S} systems are suspected to harbor a resonant chain. 

Among the observed exoplanetary systems, many planet pairs are not discovered at the exact commensurability of a MMR (e.g., with period ratio 1.5 for a 3:2 MMR), but instead depart from it. Regarding first-order MMRs, planets tend to pile up at a larger value of $n_1/n_2$ than the exact integer ratio \citep{2014ApJ...790..146F}. The deviation from the exact MMR can be measured with the \emph{resonance offset} which is defined by 
\begin{equation}
\Delta_{(p+q)/p} = P_2/P_1 - (p+q)/p
\label{eq:offset}
\end{equation}
for a system near a $(p+q)/p$ MMR. In the smooth convergent migration scenario, the spreading from exact MMR can be explained either by the damping action of the disk \citep[see, for example,][]{2014AJ....147...32G,2017A&A...602A.101R,2019A&A...625A...7P,2022MNRAS.514.3844C}, or by dissipative mechanisms such as tidal interactions with the central star once the disk has dissipated \citep{2010MNRAS.405..573P,2013AJ....145....1B,2014A&A...566A.137D}. 

\begin{figure}
    \centering
    \includegraphics[width=\columnwidth]{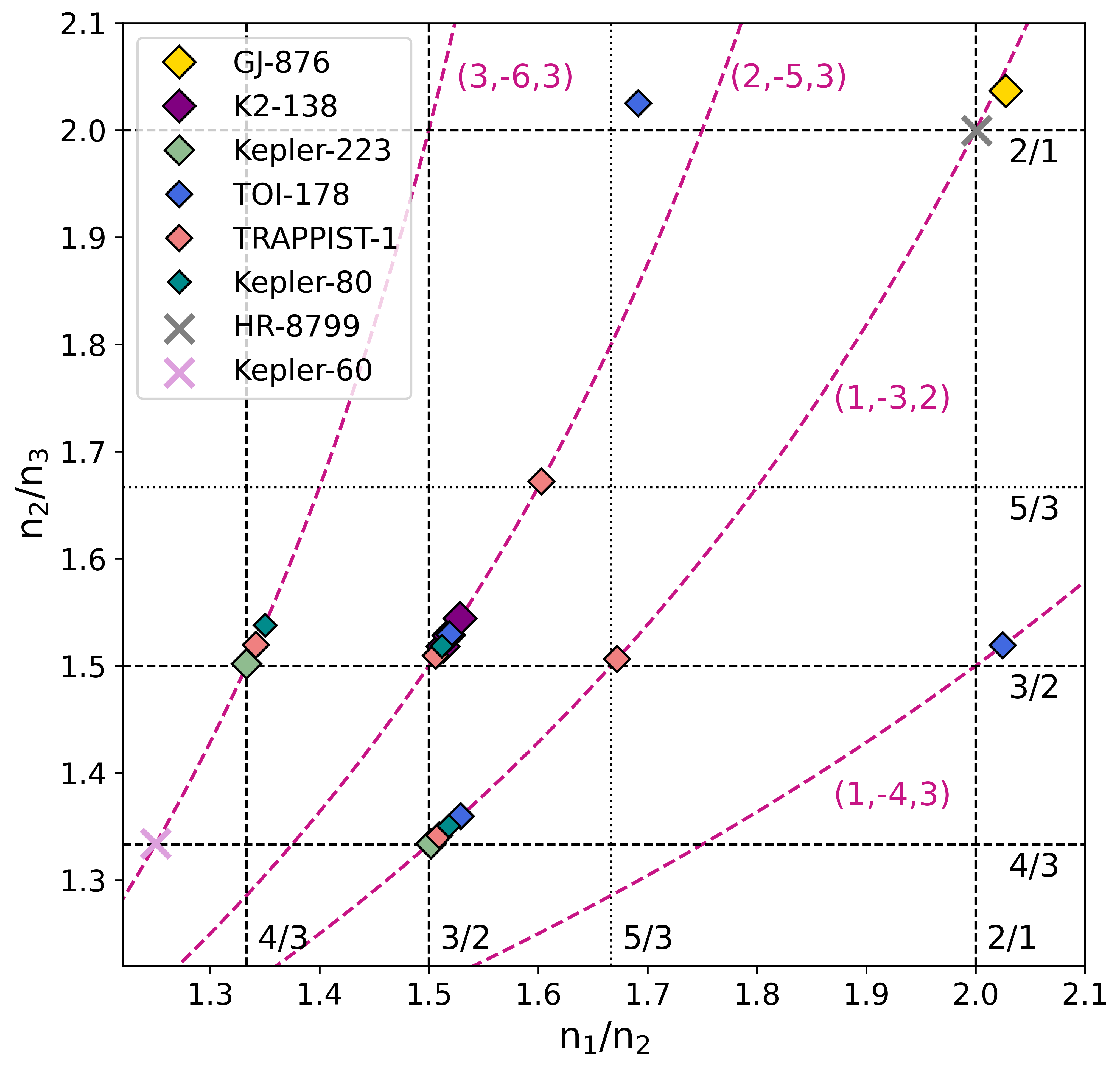}
    \caption{Departures from exact 2-body MMRs for the detected resonant chains in the $(n_1/n_2,n_2/n_3)$ plane. Vertical and horizontal lines indicate 2-planet MMRs, while the dashed purple curves indicate 3-planet resonances (the triplets refer to the coefficients $(k_1, k_2, k_3)$ of Eq.~\ref{eq:res0}). Diamonds represent the systems with confirmed resonant chains and crosses suspected resonant chains. }
    \label{fig:sys}
\end{figure}

The departures from the 2-body MMRs for the eight resonant chains currently discovered are shown in Fig.~\ref{fig:sys} in the $(n_1/n_2,n_2/n_3)$ plane, where each symbol represents a sub-system of three adjacent resonant planets. Several sub-systems are clearly shifted from the exact location of both 2-planet MMRs (i.e., intersection of the vertical and horizontal black dashed lines). However, the deviations follow a zeroth-order 3-planet resonance (also called Laplace-like resonance) represented in purple dashed curves and defined by the resonance condition
\begin{equation}
k_1 n_1 + k_2 n_2 + k_3 n_3 \simeq 0,
\label{eq:res0}
\end{equation}  
where $q_{\rm 3pl} = |k_{1} + k_{2} + k_{3}| =0$ is the order of the resonance \citep[see e.g.,][for details on the resonance structure]{2018MNRAS.477.1414C}.

The objective of this work is twofold and tackles important issues for tightly packed planetary systems. On the one hand, we aim to analyze more deeply the resonance offsets observed among the STIPs which harbor 3-planet resonances and see whether the trend observed in the offsets can be analytically predicted. On the other hand, we are interested in the formation process of resonant chains and aim to analyze how tides play an important role in shaping the resonance offsets for the STIPs, since it is well known that after the migration phase, more exactly at the dispersal of the disk, tidal interactions with the star become particularly important within the STIPs. 

It has been shown that the main effect of tides raised by the central star is to drive the planets out of resonance or even break the resonant chains, as was shown by, for example, \citet{2003ASPC..294..177N,2007ApJ...654.1110T,2012ApJ...756L..11L,2017MNRAS.470.1750I}. On the contrary, tidal dissipation would likely maintain specific sets of ratios of orbital periods, which dynamically enforces 3-body resonances \citep{2010MNRAS.405..573P,2015IJAsB..14..291P,2016MNRAS.455L.104G}. Analytical models for resonant chains were proposed by \citet{1983Icar...53...55H} for the Laplace resonance in the Galilean satellites, by \citet{2015IJAsB..14..291P} for 3-planet resonances, and by \citet{2017A&A...605A..96D} for multi-planetary resonant chains. In particular, this last work revealed the possible equilibrium configurations around which a resonant system could librate and suggested to use this information to constrain the migration scenarios. 

Different works analyzed
the role of tides in the long-term evolution of resonant chains. \cite{2021AJ....161..290S} analyzed the preferred libration centers for three-body angles through tidally damped N-body integrations. \citet{2021CeMDA.133...30P} considered the circularization due to tides from the star to describe the evolution of HD 158259 and K2-138 multi-planet systems in Laplace-like resonant chains. \citet{2021AJ....162...16G} studied the evolution of Kepler-221, for which the 3-planet resonant angles are not found librating, and favored obliquity tides to reproduce the mutually inclined architecture of Kepler-221. Since the discovery of TRAPPIST-1 system, the disc-induced migration and tidally-damped evolution of the 7-planet resonant chain was extensively studied by many authors \citep[e.g., ][]{2022MNRAS.tmp.1824B,2022MNRAS.511.3814H,2022A&A...658A.170T}, while the actual resonant evolution of the inner pair is still an open question.

Recently, \citet{2022MNRAS.513..541C} analyzed from a statistical point of view the correlation between the observed distribution of compact low-mass multi-planet systems and the 2- and 3-planet resonant structures and showed evidence in favor of a correlation with 3-planet resonances, which could be statistically explained by tides.
Note also that the survival of Laplace-like resonances with tidal effects was newly investigated by \citet{2021A&A...655A..94C} for the satellite case (where the tides act in a different manner since the planets orbit a usually slow-rotating central star while the satellites orbit a fast-rotating central planet).

While the above mentioned studies explored each system with a resonant chain individually, we offer for the first time a general study of the preservation of the resonant patterns observed in several systems, with the same general setup. The plan of the paper is as follows. A detailed offset analysis of the five confirmed STIPs with resonant chains (i.e., Kepler-80, Kepler-228, K2-138, TOI-178, and TRAPPIST-1) is conducted in Sect.~\ref{sect:obs}. In particular, we show how the observed trend in the offsets can be analytically predicted. In Sect.~\ref{sect:model}, we perform N-body simulations for the formation of these systems including both planetary orbital migration and tidal dissipation raised by the host star, and discuss the importance of tides in shaping the offset trend. Finally, Sect.~\ref{sect:discussion} is devoted to summary and conclusions.

\section{Offset analysis for systems with resonant chains}
\label{sect:obs}

In this section, we analyze the resonance offsets between adjacent pairs for five multi-planetary systems, namely Kepler-80, Kepler-223, K2-138, TOI-178, and TRAPPIST-1 whose orbital parameters are given in Table~\ref{tab:sys}. 
 
\begin{figure}
    \centering
    \includegraphics[width=\columnwidth]{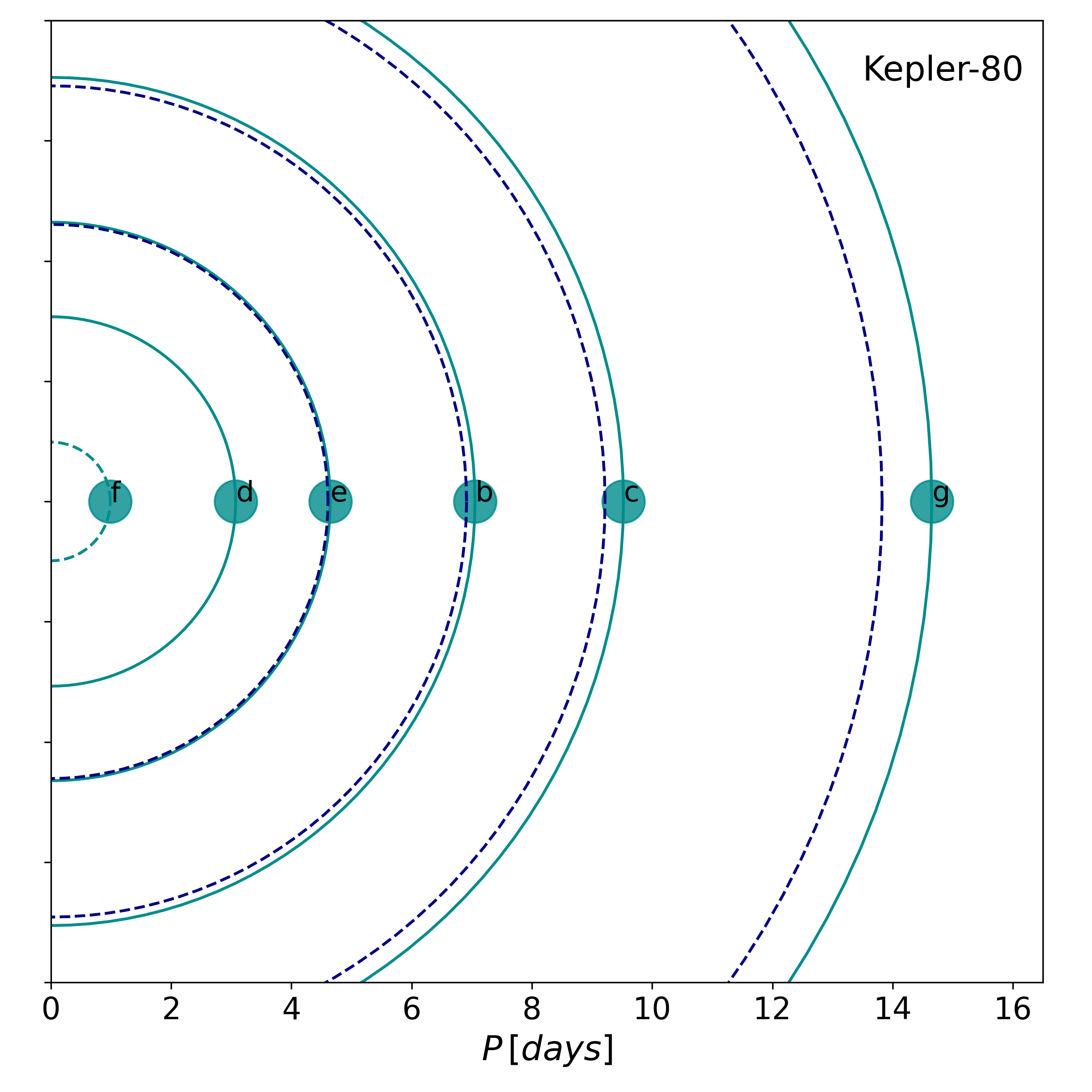}    
    \caption{Schematic representation of the Kepler-80 system. Cyan lines indicate the orbital periods from \citet{2021AJ....162..114M} and dashed blue ones the orbits at exact resonance with the preceding planet. Since the inner planet is dynamically decoupled from the resonant chain formed by the four middle planets, we start the computation of the exact resonances based on the location of Kepler-80 d. Note that the positions along the orbits are not representative, as well as the size of the points for the  planets.}
    \label{fig:Kep_80}
\end{figure}
\begin{figure}
    \centering
    \includegraphics[width=\columnwidth]{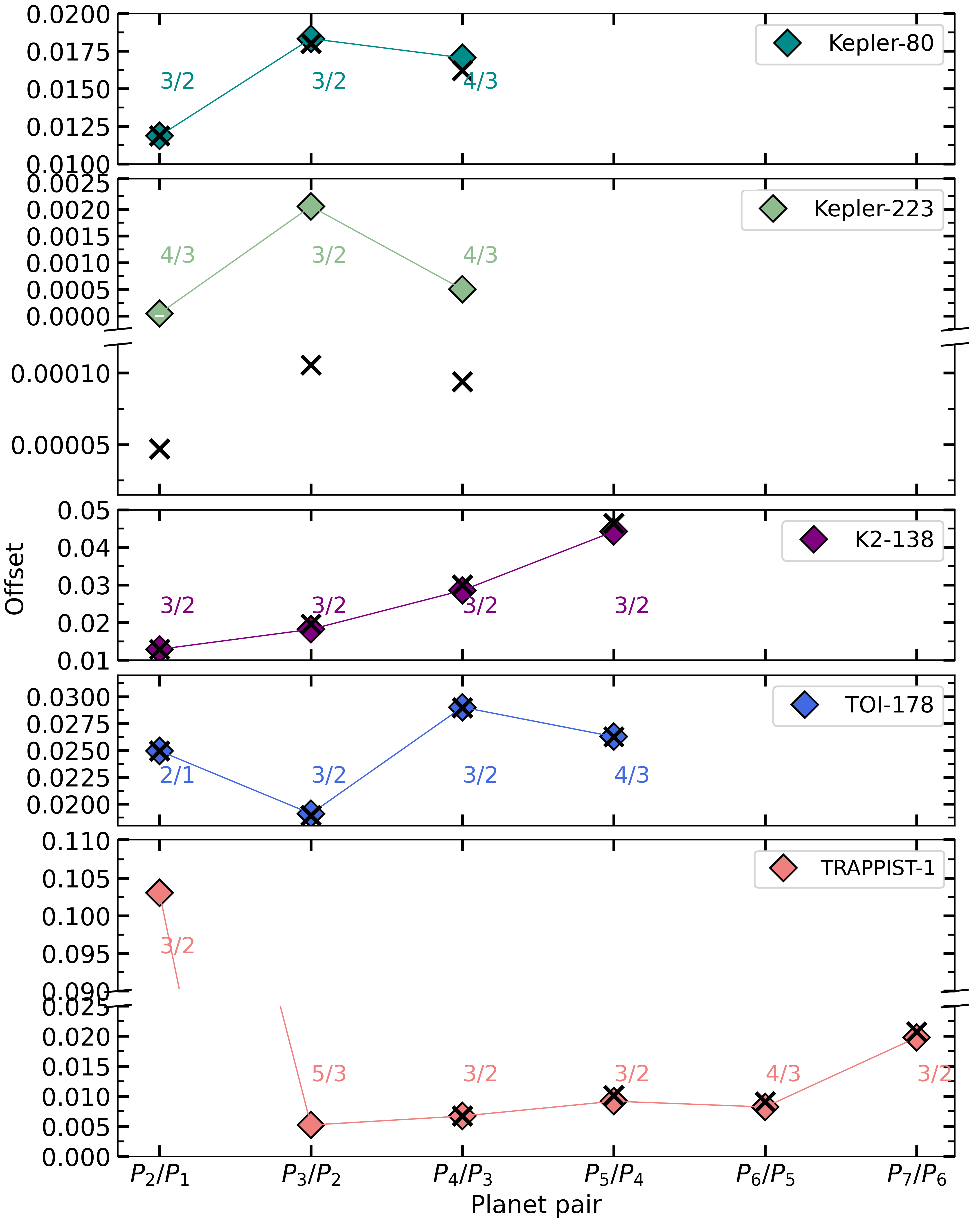}
    \caption{Resonance offsets of consecutive pairs for the five systems in resonant chain analyzed in this work. Black crosses show the offsets that the pairs should have if they were in exact 3-planet resonances as expressed in Eq.~\ref{eq:res0}.}
    \label{fig:off_systems}
\end{figure}

\begin{table*}
    \centering
    \begin{tabular}{crcccccccc}
    \hline
    & & $M_\star [{\rm M}_\odot]$ & $t_\star$ [Gyr] & $m_i \, [{\rm m}_\oplus]$ & $\rho_i \, \left[ \rho_\oplus \right]$ & $P_i$ [day] & 
    $(p+q)/p$ & $\Delta_{(p+q)/p}$ & $\tau^{\rm tid}_{e_i}$ [yr] \\
    \hline
    & f & & &  --  &  --  & 0.98678 &                       &                          & -- \\
    & d & & & 5.95 & 14.6 & 3.07222 & \multirow{-2}{*}{3/1} & \multirow{-2}{*}{0.1133} &  $8.8 \times 10^4$ \\
    & e & & & 2.97 & 6.9  & 4.64489 & \multirow{-2}{*}{3/2} & \multirow{-2}{*}{0.0118} &  $2.4 \times 10^5$ \\
    & b & & & 3.5  & 1.45 & 7.05246 & \multirow{-2}{*}{3/2} & \multirow{-2}{*}{0.0186} &  $9.7 \times 10^4$ \\
    & c & & & 3.49 & 1.22 & 9.52355 & \multirow{-2}{*}{4/3} & \multirow{-2}{*}{0.0168} &  $3.2 \times 10^5$ \\
    \multirow{-6}{*}{Kepler-80} & g & \multirow{-6}{*}{0.73} & \multirow{-6}{*}{$2 \pm 1$} & 0.065 & -- & 14.6455 & \multirow{-2}{*}{3/2} & \multirow{-2}{*}{0.0385} &  -- \\
    \hline
    & b & & & 7.4 & 0.279 & 7.38449  &  &    & $1 \times 10^5$\\
    & c & & & 5.1 & 0.129 & 9.84564  & \multirow{-2}{*}{4/3} & \multirow{-2}{*}{0.0000} & $1.3 \times 10^5$ \\
    & d & & & 8   & 0.056 & 14.78869 & \multirow{-2}{*}{3/2} & \multirow{-2}{*}{0.0021} & $1.4 \times 10^5$ \\
     \multirow{-4}{*}{Kepler-223} & e & \multirow{-4}{*}{1.125} & \multirow{-4}{*}{$\sim 6 $} & 4.8 & 0.051 & 19.72567 & \multirow{-2}{*}{4/3} & \multirow{-2}{*}{0.0005} & $5.9 \times 10^5$ \\ 
    \hline
    & b & & & 3.1 & 0.89 & 2.3531  &  &    & $8.1 \times 10^3$ \\
    & c & & & 6.3 & 0.51 & 3.56    & \multirow{-2}{*}{3/2} & \multirow{-2}{*}{0.0129} & $1.2 \times 10^4$ \\
    & d & & & 7.9 & 0.58 & 5.4048  & \multirow{-2}{*}{3/2} & \multirow{-2}{*}{0.0182} & $7.8 \times 10^4$ \\
    & e & & & 13 & 0.33 & 8.2615  & \multirow{-2}{*}{3/2} & \multirow{-2}{*}{0.0285} & $1.4 \times 10^5$ \\ 
    \multirow{-5}{*}{K2-138} & f & \multirow{-5}{*}{0.93} & \multirow{-5}{*}{$2.3^{+0.44}_{-0.36}$} & 8.7 & 0.38 & 12.7576 & \multirow{-2}{*}{3/2} & \multirow{-2}{*}{0.0442} & $1.5 \times 10^6$ \\ 
    \hline
    & b & & & 1.5  & 0.98 & 1.914558 &  &    & $4.9 \times 10^4$ \\
    & c & & & 4.77 & 1.02 & 3.23845 & \multirow{-2}{*}{5/3} & \multirow{-2}{*}{0.0248} &  $2.4 \times 10^4$ \\
    & d & & & 3.01 & 0.177 & 6.5577 & \multirow{-2}{*}{2/1} & \multirow{-2}{*}{0.0250} & $3.7 \times 10^4$ \\
    & e & & & 3.86 & 0.360 & 9.961881 & \multirow{-2}{*}{3/2} & \multirow{-2}{*}{0.0191} & $6.3 \times 10^5$ \\ 
    & f & & & 7.72 & 0.65 & 15.231915 & \multirow{-2}{*}{3/2} & \multirow{-2}{*}{0.0290} & $6.7 \times 10^6$ \\
    \multirow{-6}{*}{TOI-178} & g & \multirow{-6}{*}{0.65} & \multirow{-6}{*}{$7.1^{+6.1}_{-5.3}$} & 3.94 & 0.166 & 20.7095 & \multirow{-2}{*}{4/3} & \multirow{-2}{*}{0.0263} & $4.1 \times 10^6$ \\
    \hline
    & b &  & & 1.374 & 0.987 & 1.510826 &  &    & $5.1 \times 10^2$ \\
    & c &  & & 1.308 & 0.991 & 2.421937 & \multirow{-2}{*}{3/2} & \multirow{-2}{*}{0.1031} & $4.1 \times 10^3$ \\
    & d &  & & 0.388 & 0.792 & 4.049219 & \multirow{-2}{*}{5/3} & \multirow{-2}{*}{0.0052} & $5.9 \times 10^4$ \\
    & e &  & & 0.692 & 0.889 & 6.101013 & \multirow{-2}{*}{3/2} & \multirow{-2}{*}{0.0067} & $2.9 \times 10^5$ \\ 
    & f &  & & 1.039 & 0.911 & 9.207540 & \multirow{-2}{*}{3/2} & \multirow{-2}{*}{0.0092} & $1.3 \times 10^6$ \\
    & g &  & & 1.321 & 0.917 & 12.352446 & \multirow{-2}{*}{4/3} & \multirow{-2}{*}{0.0082} & $4.2 \times 10^6$ \\ 
    \multirow{-7}{*}{TRAPPIST-1} & h & \multirow{-7}{*}{0.898} & \multirow{-7}{*}{$7.6 \pm 2.2$} & 0.326 & 0.755 & 18.772866 & \multirow{-2}{*}{3/2} & \multirow{-2}{*}{0.0198} & $4.7 \times 10^7$ \\ 
    \hline
    \end{tabular}
    \caption{Parameters of the five systems in resonant chain analyzed in this work. The columns successively give for each system the star mass and age, planetary masses and densities, orbital periods, 2-planet MMRs between adjacent planets, resonance offsets, and $e$-damping timescales due to tides raised by the star using $Q'=0.05$. Data for Kepler-80 was taken from \citet{2016AJ....152..105M,2021AJ....162..114M}, for Kepler-223 from \citet{2016Natur.533..509M}, for K2-138 from \citet{2019A&A...631A..90L}, for TOI-178 from \citet{2021A&A...649A..26L}, and for TRAPPIST-1 from \citet{Burgasser_2017,2021PSJ.....2....1A}.}
    \label{tab:sys}
\end{table*}

\subsection{Resonance offsets for Kepler-80, Kepler-223, K2-138, TOI-178, and TRAPPIST-1}

We begin our analysis with Kepler-80, a 6-planet STIP orbiting a M0/K5 star with $M_\star = 0.73 \, {\rm M}_\odot$, and star age $t_\star \sim 2$ Gyr \citep{2016AJ....152..105M}. The parameters shown in Table~\ref{tab:sys} are taken from \citet{2021AJ....162..114M}. The six planets of Kepler-80 belong to a resonant chain, in the sense that all adjacent pairs are found near a 2-body MMR, as indicated in the sixth column of Table~\ref{tab:sys}. Like many STIPs, the 2-body MMRs are more often first-order MMRs. By looking at the transit timing variations (TTVs), \citet{2016AJ....152..105M,2021AJ....162..114M} concluded that the four planets Kepler-80 d, e, b, and c dynamically interact in a resonant chain, while the innermost planet Kepler-80 f is decoupled from the rest of the planets. Moreover, the libration of resonant angles involving Kepler-80 g is not guaranteed. As a result, we only investigate in the following the four-body resonant chain between the middle planets Kepler-80 d, e, b, and c. In Fig.~\ref{fig:Kep_80}, we display a schematic view of the system. The solid cyan lines represent the orbits of the different planets with the orbital periods given by the observations (assuming them circular), while the dashed blue lines represent the orbits that planets should have if they were located in the exact period ratio resonant chain (i.e., only planet d is considered at the observed location, the planets exterior to it being all in exact MMR in pairs). From this plot we can see that the observed periods are shifted from the exact resonant chain, with the deviation of each orbital period to the resonant one increasing with the distance to the star. It is important to stress that this view does not depict the usual quantity for the deviation from the exact MMR known as resonance offset (see Eq.~\ref{eq:offset}).

The resonance offsets (defined as in Eq.~\ref{eq:offset}) for the four middle planets of Kepler-80, whose planet pairs are successively close to the 3/2, 3/2, and 4/3 MMRs, are shown in the top panel of Fig.~\ref{fig:off_systems} (colored diamonds). While the deviation from the exact resonant period increases from the innermost to the outermost planet in Fig.~\ref{fig:Kep_80}, it becomes clear that the resonance offsets do not increase monotonically with the distance to the star. The offset of the second pair increases with respect to the offset of the first pair, while the offset of the last pair decreases with respect to the one of the second pair. 

The resonance offsets are also displayed in Fig.~\ref{fig:off_systems} for Kepler-223, K2-138, TOI-178, and TRAPPIST-1, as well as in the column labeled as $\Delta_{(p+q)/p}$ of Table~\ref{tab:sys}. Kepler-223 is a 4 sub-Neptune planetary system orbiting a Sun-like star ($M_\star= 1.25 \, {\rm M}_\odot$ and $t_\star \sim 6$ Gyr), for which TTVs, and thus well constrained masses, were reported for the four planets by \citet{2016Natur.533..509M}. The consecutive pairs of planets are close to the 4/3, 3/2, and 4/3 MMRs, with almost negligible departure from exact resonance. The long-term stability of this resonant compact system and its possible migration history were confirmed by \citet{2016Natur.533..509M}.

K2-138 system consists of an early K-type star ($M_\star= 0.93 \, {\rm M}_\odot$ and $t_\star \sim 2.3$ Gyr) and six planets, five of which were found in the longest chain of 3:2 MMRs \citep{2018AJ....155...57C,2019A&A...631A..90L}, the outermost planet K2-138 g being most likely dynamically decoupled from the resonant chain. It has been shown that the specific chain of 3/2 MMRs is the usual outcome of N-body simulations \citep{2022arXiv220112687M}. Recently, with a semi-analytic model including orbital circularization induced by tidal interaction with the central star, \citet{2021CeMDA.133...30P} highlighted the importance of the interlinked 3-planet resonances on the long-term evolution of the system.

TOI-178 system was first identified as a possible co-orbital system \citep{2019A&A...624A..46L} with two of the planets at roughly the same orbital period. Further observations with CHEOPS, ESPRESSO, NGTS, and SPECULOOS showed instead that the system consists in 6 planets interlinked by resonant chains \citep{2021A&A...649A..26L}.
The consecutive pairs of planets are close to the 5/3, 2/1, 3/2, 3/2, and 4/3 MMRs. This resonant chain is also characterized by 3-planet resonances between consecutive triplets for all planets except the innermost one, which seems to greatly stabilize the system. As in the case of Kepler-80,  we decided to remove the innermost planet from our analysis since this planet is not involved in any Laplace resonance (note that there is no highlighted 3-planet MMR for this triplet in Fig.~\ref{fig:sys}). 
The mass of the TOI-178 star is $M_\star= 0.65 \, {\rm M}_\odot$, and the star age is estimated with great uncertainty to $t_\star \sim 7.1^{+6.1}_{-5.3}$ Gyr \citep{2021A&A...649A..26L}.

TRAPPIST-1 \citep{2016Natur.533..221G,2017Natur.542..456G} consists of seven Earth-sized planets orbiting an ultra-cool dwarf star with mass $M_\star = 0.0898 \, {\rm M}_\odot$ and $t_\star = 7.6 \pm 2.2$ Gyr \citep{Burgasser_2017}. The planets are in interlinked 2-planet MMRs and 3-planet resonances, with period ratios close to 8/5, 5/3, 3/2, 3/2, 4/3, and 3/2 \citep{2017NatAs...1E.129L}. The TRAPPIST-1 system presents strong TTVs, allowing for an improvement of the mass estimates of the seven planets \citep[see][]{2021PSJ.....2....1A}. With the recently refined best-fit solution, the resonant dynamics of the system was analyzed by \citet{2022A&A...658A.170T} who emphasized the peculiar resonant dynamics of the innermost
pair of planets and cast doubt on their formation and evolution in the 8/5 MMR. They also confirmed the possible formation, by disc-induced migration, of a chain of intricate resonances which accurately reproduces the observed TTVs, when adopting particular disc conditions.

The four extrasolar systems described above exhibit different trends in the offsets, as shown in Fig.~\ref{fig:off_systems}. For the 4-planet Kepler-223 system, the offsets first increase and then decrease, similarly to Kepler-80. The offsets of the 5-planet K2-138 system monotonically increase. The 6-planet TOI-178 system presents successive increases and decreases of the resonance offsets. Another complex behavior of the resonance offsets is observed for the 7-planet TRAPPIST-1 system. The origin of the trends observed in the offsets will be discussed in the next section.

\subsection{Analytical estimation of the offsets induced by 3-planet resonances}
\label{sub:estimation}

In this section we show that the trends in the resonance offsets observed in Fig.~\ref{fig:off_systems} are linked with 3-planet resonances. To do so, we add in Fig.~\ref{fig:off_systems} with black crosses, the resonance offsets that the pairs should have if they were in exact 3-planet resonances. More precisely, assuming that the first two planets of the resonant chain are in the observed location (i.e., the first black cross coincides with the first diamond in Fig.~\ref{fig:off_systems}), we computed the period of the third planet guided by the 3-planet resonance relation. If the 3-planet resonance is a zeroth-order one ($q_{\rm 3pl}=0$), Eq.~\ref{eq:res0} gives 
\begin{equation}
    P_3 = \frac{P_2}{1-\frac{k_1}{k_3}\left(\frac{P_2}{P_1}-1\right)},
    \label{eq:P3_zero}
\end{equation} while if the triplet is in a first-order 3-planet resonance ($q_{\rm 3pl}=1$), we obtain
\begin{equation}
    P_3 = \frac{P_2}{1-\frac{1}{k_3} -\frac{k_1}{k_3}\left(\frac{P_2}{P_1}-1\right)}
    \label{eq:P3_one}
\end{equation}
\citep[see also the analysis of first-order 3-planet resonances of][]{2021CeMDA.133...39P}. With this value, we calculated the resonant offset of the second pair (whose planets are thus located in exact 3-planet resonance with the inner one). For the next triplet, we used the observed period of the second planet and the computed period of the third one and repeated the procedure, and so on. The results for the multi-planetary systems Kepler-223, Kepler-80, K2-138, TOI-178, and TRAPPIST-1 are shown in Fig.~\ref{fig:off_systems} and discussed in the following.

Regarding the 4-planet Kepler-80 system, planets d-e-b are in a $(2,-5,3)$ 3-planet resonance while the second triplet e-b-c is in a $(1,-3,2)$ resonance, both triplets being thus in a zeroth-order resonance. From this plot, we clearly see that the 3-planet resonances are guiding the dynamics of Kepler-80 since the trend in the observed resonance offsets perfectly fits the analytic estimation. The same conclusion can be drawn for the 5-planet K2-138 system, with all triplets in zeroth-order resonance. For Kepler-223, the analytical estimate does not coincide with the observations. Although similar offset increase and then decrease are observed, there is a difference of two orders of magnitude in the offset values. Since the innermost planet of the 5-planet TOI-178 is not part of the resonant chain, we begin the computation of the 3-planet resonant chain from the second planet. Again, we observe a perfect matching between the observed and analytically estimated values, showing similar successive increases and decreases. Finally, the complex behavior of the resonant offsets of TRAPPIST-1 is also well reproduced with the analytical estimates for the well-known resonant chain of the five outer planets (more details are provided in Sect.~\ref{sect:rest}).

In the next section, the formation of the resonant chains will be investigated by means of N-body simulations including both orbital migration and tidal dissipation from the host star. We aim to give insights into the role of tides within the STIPs with resonant chains, in particular on the reproduction of the observed trends of the resonant offsets. 

\begin{figure*}
    \centering
    \includegraphics[width=1.5\columnwidth]{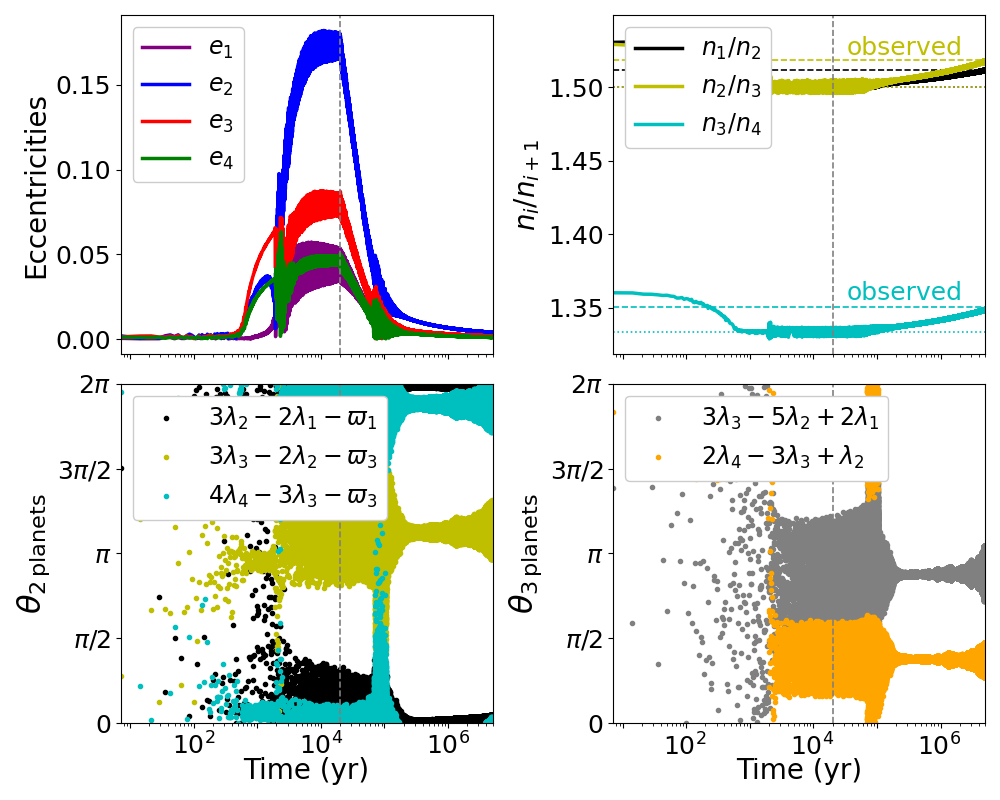}
    \caption{Typical time evolution for the four planets of the resonant chain of Kepler-80 (d, e, b, and c) with $Q'=0.05$ and physical parameters from Table~\ref{tab:sys}. In the top left panel the eccentricities, in the top right panel the mean-motion ratios of adjacent planets, in the bottom left panel the two-body resonant angles and in the bottom right panel the three-body resonant angles. }
    \label{fig:evolution}
\end{figure*}

\section{Effect of tides on the resonance offsets} 
\label{sect:model} 

In this section we focus on the late-stage formation of planetary systems with resonant chains, in particular the final stage of the protoplanetary disk phase when the systems achieve the configurations we observe nowadays. More precisely, we study the evolution of the systems during their migration in the gas disk within which captures in MMRs take place and after the dispersal of the disk when tidal interactions with the star influence significantly the long-term evolution of close-in planets. The simulations realized in this work are described in Section~\ref{sec:code}. In Section~\ref{sect:k80}, a discussion on the tidal parameters is given. Applications to the five multi-planetary systems analyzed in the previous section are achieved in Section~\ref{sect:rest}.

\subsection{Code and set-up of the simulations}
\label{sec:code}

To perform simulations of the late-stage formation and long-term evolution of planetary systems with resonant chains, we used the \textsc{REBOUNDx} N-body code \citep{2020MNRAS.491.2885T} with the WHFast symplectic integrator \citep{1991AJ....102.1528W,2015MNRAS.452..376R}. Both disk-induced planetary migration and tidal interactions with the host star are included. Here we do not consider the tides raised on the star by the planets, as planets of a few Earth masses will not produce an appreciable effect on the star. Considering the planetary migration, we adopted a simple prescription with constant damping timescales acting on the outer planet only. For the radial decay we considered a damping timescale of $\tau_a = 10^7$ days and for the circularization timescale $\tau_e = 10^5$ days\footnote{We also run some tests with the prescriptions given in \citet{2008A&A...482..677C} for the migration of the outer planet. As discussed in \citet{2022MNRAS.514.3844C}, when the planets evolve in the MMR during the disk phase, the eccentricity evolution continuously provides a feedback on the damping time keeping the values of the eccentricities bounded during the migration. With this more complex prescription, the results detailed here for the long term evolution of the five systems remain mostly unchanged, which underlines that the migration history is erased by the action of the tides.}. 

We fixed the initial conditions for the planets similarly to \citet{2016AJ....152..105M}: the first planet is moved outwards arbitrarily while the rest of the planets are initially located within 3\% of the nominal resonance with the preceding planet, that is, slightly more spread from MMRs than the observed systems. This also means that a very short migration phase ($\sim 2 \times 10^4$ years) is required for the systems to settle in resonant configurations. For all the simulations, the planets are assumed to be coplanar with eccentricities initially fixed to $10^{-4}$ and random orbital angles in $[0,2\pi]$.

Regarding the tides raised by the star, we used a tidal damping timescale on the eccentricity given by \citet{1966Icar....5..375G, 2021CeMDA.133...30P}\footnote{Eq.~\ref{eq:tidal} is valid when the rotation period of the central star is longer than the orbital period of the planet, which is expected for planets with orbital period of less than $\sim$ 15 days.}
\begin{equation}
    \tau_{e_i}^{\rm tid} = 7.63 \times 10^{5} \left(\frac{a_i}{0.05{\rm au}}\right)^{13/2}
    \left(\frac{M_\odot}{M_\star}\right)^{3/2} \left(\frac{m_\oplus}{m_i}\right)^{2/3}
    \left(\frac{\rho_i}{\rho_\oplus}\right)^{5/3} Q' {\rm yr},
    \label{eq:tidal}
\end{equation} 
with $\rho_i$ the mean density of the planet, $Q'=3Q/(2k_2)$ where $Q$ represents the tidal quality factor and $k_2$ the Love number. It is important to note that these parameters are generally not known for the detected exoplanets. Alternatively, the values for the terrestrial planets in the Solar System can be extrapolated and it is reasonable to use $k_2=0.3$ and $Q\in [10,500]$, allowing for $Q'$ to vary between 50 and 2500. As an example, for an Earth-mass planet at $0.05 \, {\rm au}$ and $Q'=50$, the circularization timescale due to the star is $\tau_{e}^{\rm tid} \sim 3.5 \times 10^7 \, {\rm yr}$, which implies strong tidal effects for close-in extrasolar planets in the Earth-mass regime. Higher values of $Q'$ increase the circularization timescale, and thus ask for time-consuming integrations to study the long-term evolution of the systems. For this reason, small values of $Q'$ between 0.005 and 2.5 are generally adopted in the literature in order to carry out numerical integrations over reasonable computation times \citep[see also][]{2021CeMDA.133...30P}. Although not realistic in regard to the terrestrial planets of the Solar system, these values are in agreement with the range used for TRAPPIST-1 \citep{2020A&A...644A.165B,2022MNRAS.tmp.1824B}. Let us stress that the above-mentioned values of $Q'$ are such that eccentricity damping from the disk is faster than the damping raised by the star, which means that the tidal effects from the star act on longer timescales. As in \citet{2021AJ....161..290S}, we assume that the value of $Q'$ is the same among all the planets in the system.

From Eq.~\eqref{eq:tidal} it is clear that the innermost planets are the most affected by tidal effects from the star. \citet{2013ApJ...778....7B} estimated that a super-Earth with an orbital period greater than $\sim$~10 days should experience negligible tidal effects. Although the outer planets of the systems with resonant chains studied in this work have generally periods longer than 10 days and should be less affected by the tides raised from the star, we clearly see in Fig.~\ref{fig:off_systems} that they are shifted from the exact resonant positions, pushed further away due to the resonant interactions with the inner planets for which tidal effects are important, as it will be highlighted in the next subsection.

\subsection{Discussion on the tidal parameters}
\label{sect:k80}

\begin{figure}
    \centering
    \includegraphics[width=\columnwidth]{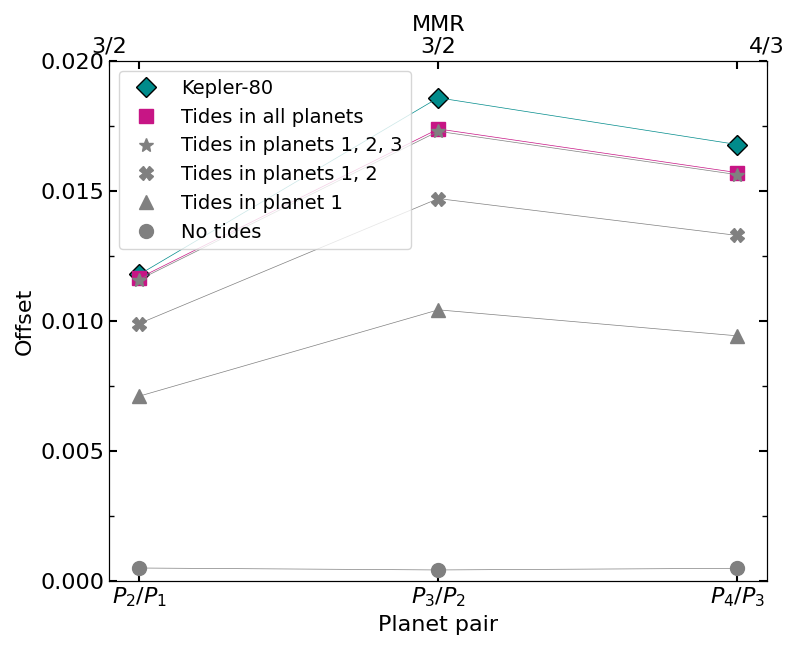}
    \caption{Resonance offsets of the consecutive pairs for the resonant chain of Kepler-80 (planets d, e, b, and c). Cyan diamonds show the observed values (as in Fig.~\ref{fig:off_systems}), while gray symbols show the simulation outcomes where no tides are considered (circles), or with tides (following Eq. \eqref{eq:tidal}) applied to the inner planet (triangles), first two planets (crosses), first three planets (stars), and all the planets (squares, see the long-term evolution in Fig.~\ref{fig:evolution}). Integrations were made for $5 \times 10^6 \, {\rm yr}$ with $Q'=0.05$ and the parameters of Table~\ref{tab:sys}. }
    \label{fig:off_methods}
\end{figure}

\begin{figure}
\centering
    \includegraphics[width=\columnwidth]{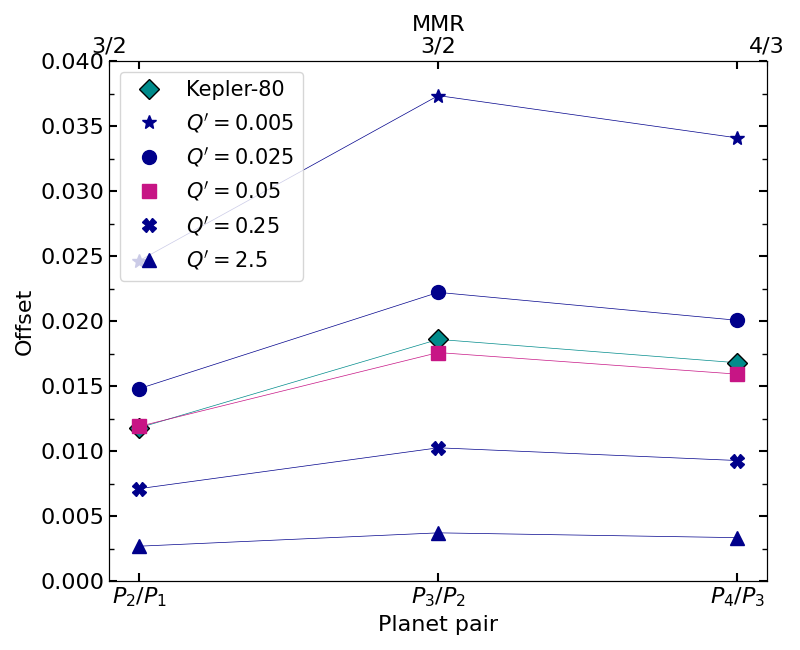}
    \includegraphics[width=\columnwidth]{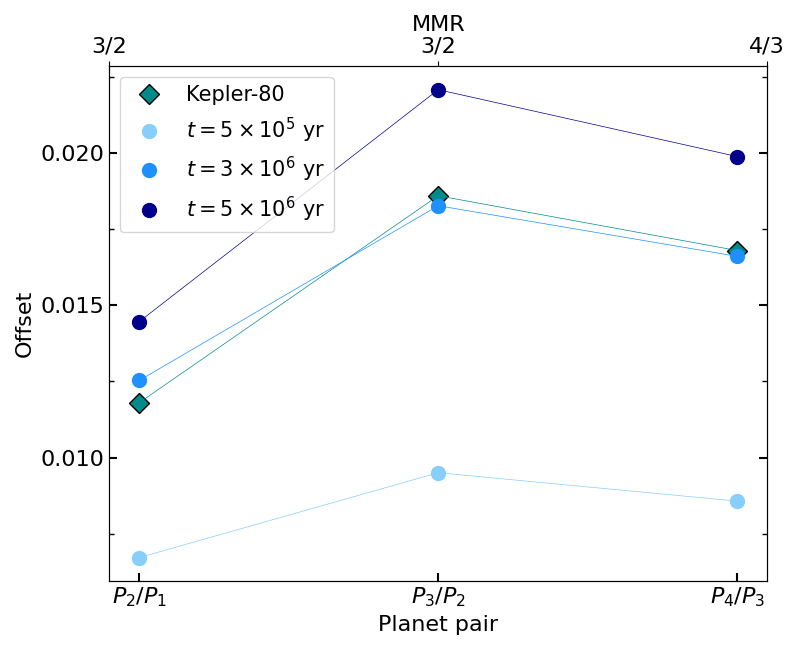}
    \caption{Resonance offsets for Kepler-80 as in Fig.~\ref{fig:off_methods} for different $Q'$ values (with simulation timescale fixed to $5 \times 10^6$ yr) in the top panel and different simulation timescales (with $Q'=0.025$) in the bottom panel. See text for more details.}
    \label{fig:off_m5}
\end{figure}

\begin{figure}
    \centering
    \includegraphics[width=\columnwidth]{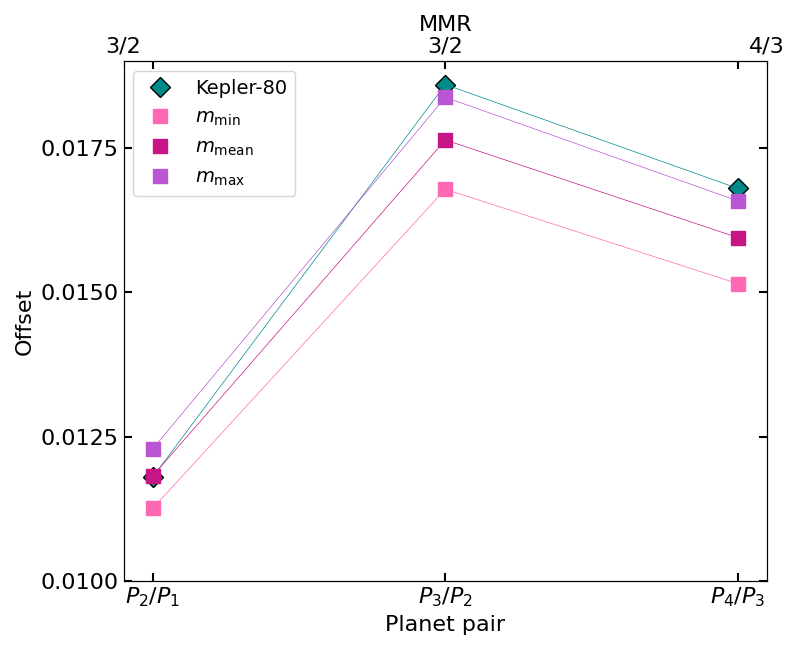}
    \caption{Resonance offsets for Kepler-80 for the lower, mean and bigger masses reported in \citet{2021AJ....162..114M} using $Q'=0.05$. } 
    \label{fig:off_Q}
\end{figure}

In order to examine the patterns arising from tidal dissipation in resonant chains, we individually investigate various parameters that could impact the outcomes. Specifically, we compare these parameters with the characteristics observed in the Kepler-80 \as{system}.
Additionally, we expand our analysis to encompass Kepler-223, \cc{K2-138,} TOI-178, and TRAPPIST-1 in the subsequent subsection.

As already pointed out in the previous section, Kepler-80 hosts a resonant chain made of four planets (Kepler-80 d, e, b, and c) near the 3/2, 3/2, 4/3 MMRs, respectively \citep{2016AJ....152..105M,2021AJ....162..114M}.  
In Fig.~\ref{fig:evolution} we show a typical time evolution for the 4-planet resonant chain of Kepler-80. When the planets pairs are trapped in MMRs (see the 2-body resonant angles in the bottom left panel and the 3-body resonant angles in the bottom right panel) during their disk-induced migration, the eccentricities begin to increase to settle around fixed values. These values then decrease due to the action of the tides raised by the star at the disk dispersal (dashed vertical line) and reach values close to the estimated values given by the observations of Kepler-80. Simultaneously, the planet pairs stop their oscillations around the exact commensurability (horizontal dotted line in the top right panel) and tend to the observed departure from MMR (horizontal dashed line). We clearly see that the orbital circularization induced by the tides with the host star is associated to the resonance offsets observed in Fig.~\ref{fig:off_systems}. Let us note that shifts of the libration centers for the resonant angles can be observed due to the effect of the tides (see the bottom plots), as already noticed in \citet{2021AJ....161..290S}. For this simulation, we considered $Q'=0.05$, initial orbital periods of $P_i = (3.7, 5.66, 8.66, 11.78)$, and the planetary masses and densities from Table~\ref{tab:sys}. 

To further analyze the effects of tides on the long-term evolution of the resonant chain, we display in Fig.~\ref{fig:off_methods} the resonance offsets of the three planet pairs analyzed for Kepler-80, as found at the end of the simulation of Fig.~\ref{fig:evolution}, namely $5 \times 10^6 \, {\rm yr}$ (pink square symbols). The observed resonant offsets are shown with cyan diamonds, as they will be for the rest of the section. We clearly see that the trend observed in the offsets for Kepler-80 in Section~\ref{sect:obs} and the trend in the offsets resulting from our simulation involving planetary migration and tidal interactions with the star are very similar to each other. The importance of the tidal effects on the resonance offsets is emphasized by the results of additional simulations shown in Fig.~\ref{fig:off_methods} where we did not include the tidal effects on all the planets on the resonant chain. In particular, the resonance offsets are shown with gray circles when no tides are considered in the simulation, gray triangles when the tides are only applied to the inner planet of the resonant chain (planet d), gray crosses when they are applied on planets d and e and gray stars when they are applied on the third planet as well (planet b). 

When no tides are considered on the system, the offsets are not significant at all and all the planets depart similarly from the nominal resonances. Let us recall that we considered here very simple migration prescription (for instance, the inner edge of the disk and the time decrease of the disk mass are not included here) and many works properly focus on the disk-induced resonance offsets with more sophisticated prescriptions \citep[see e.g., ][]{2019A&A...625A...7P,2020A&A...643A..11T,2022MNRAS.514.3844C} to which we refer to for more details. More importantly, when the tides from the star act on the system, the trend in the offsets is always similar to the one given by the observations, whatever the number of planets on which the tides are applied. The outcomes for which tides are applied on three and four planets of the resonant chain are nearly superimposed in Fig.~\ref{fig:off_methods}, reinforcing the idea that tides are more important for the inner planets. This does not mean that the outermost planet is not affected by tides, but the timescale of $5 \times 10^6 {\rm yr}$ is not long enough to see different amplitudes in the resonance offsets. Indeed, by looking at Eq.~\ref{eq:tidal}, we see that the time needed for the innermost planet to circularize its orbit due to tides is $\tau_{e_1}^{\rm tid} \sim 8.8 \times 10^4 \, {\rm yr}$ (i.e., nearly the integration timescale of Fig.~\ref{fig:evolution}), while the tidal damping timescale increases rapidly for more distant planets, namely $\tau_{e_2}^{\rm tid} \sim 2.4 \times 10^5$,  $\tau_{e_3}^{\rm tid} \sim 9.7 \times 10^4$, and $\tau_{e_4}^{\rm tid} \sim 3.2 \times 10^5 \, {\rm yr}$. In the following we will always consider the tides raised by the star on all the planets. 

All the previous investigations were realized with $Q'=0.05$ in order to be able to reproduce the Kepler-80 evolution to the observed values in $5 \times 10^6$ yr. The impact of the choice of the $Q'$ value on the previous results is analyzed in Fig.~\ref{fig:off_m5} where the top panel shows the resonance offsets obtained with different $Q'$ values in the simulations. Four additional values are considered, $Q'=0.005$, $0.025, 0.25$ and $2.5$, while the timescale of the simulations remains unchanged (i.e., $5 \, \times 10^6 \, {\rm yr}$). As it can be observed, all the $Q'$ values give the same trend in the offsets (i.e., the offset of the second pair is bigger than the one of the first pair, while the last one presents intermediate offset values), only the offset amplitudes vary. As explained previously, this amplitude variation is due to the timescale adopted for the simulations, as illustrated in the bottom panel of Fig.~\ref{fig:off_m5}. The resonance offsets for Kepler-80 with the tidal parameter fixed to $Q'=0.025$ are displayed for three different integration times: $5 \, \times 10^5 \, {\rm yr}$, $3 \, \times 10^6 \, {\rm yr}$, and $5 \, \times 10^6 \, {\rm yr}$. As expected from Eq.~\ref{eq:tidal}, we see that simulations with lower $Q'$ values can reproduce the observed resonance offsets of Kepler-80 in a smaller timescale, namely $\sim 3 \, \times 10^6 \, {\rm yr}$ for $Q'=0.025$. However, it is not clear when tides stop acting, or even if there is a plateau in the variation of the orbital elements. It is most likely that unless other effect is present, the dissipation in the system leading to orbital spreading will be active and accumulating throughout the lifetime of the system. Thus, the amplitude in the offsets will increase with the integration time span.
Finally, we analyze the sensitivity of the resonance offsets to the errors in the planetary masses. To do so, we simulated the smaller and bigger masses reported in \citet{2021AJ....162..114M} with a fixed value of $Q'=0.05$ (where it is fair to assume that the smaller/bigger masses are associated to the lower/higher densities reported). 
The offset values reached after an integration time span of $5 \times 10^6 \, {\rm yr}$ are shown in Fig.~\ref{fig:off_Q}. Again, although the amplitude of the offsets change with the individual masses, the trend in the offsets is preserved, and in all cases the 2 and 3 body resonant angles are found librating. 
While Fig.~\ref{fig:off_Q} suggests that considering a larger mass yields a better reproduction of the system's offset, as explained in Fig.~\ref{fig:off_m5}, the amplitude of the offset will continue to increase as the system ages. The key finding is that, regardless of the varying masses, we consistently observe the same outcome: the offset of the second pair is greater than that of the first pair, while the offset of the last pair is smaller than the preceding one. This pattern holds true across different $Q'$ values.

From the previous analysis, we can highlight that the tidal effects are a robust mechanism able to reproduce the trend of the resonance offsets observed for Kepler-80, while the system evolves guided by the 3-planet dynamics. Since we are interested here by the trend of the offsets, we will fix $Q'=0.05$ in the following and keep in mind that the amplitude in the offsets could be reproduced by fine-tuning the simulation timescale. Moreover, we have shown that the trend in the offsets is not influenced by the uncertainties of individual masses.

\subsection{Applications to observed systems}
\label{sect:rest}
\begin{figure}
    \centering
    \includegraphics[width=\columnwidth]{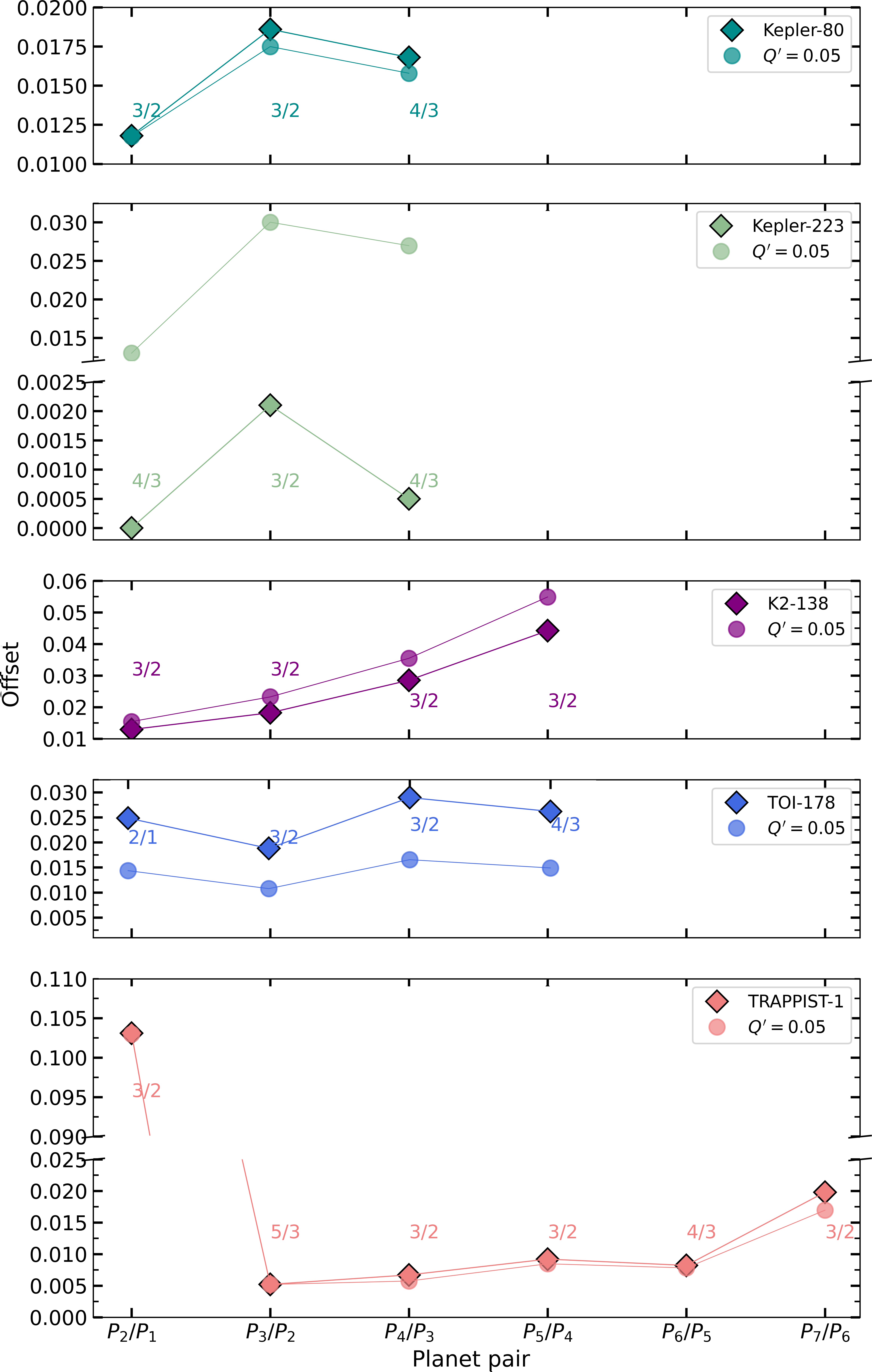}
    \caption{Resonance offsets of consecutive pairs for the five systems in resonant chain analyzed in this work, as in Fig.~\ref{fig:off_systems} (diamonds). Dots show the outcomes of the simulations performed with the tidal factor $Q'=0.05$ for $5 \times 10^6$ yr. }
    \label{fig:off_sys_sims}
\end{figure}
\begin{figure}
    \centering
    \includegraphics[width=\columnwidth]{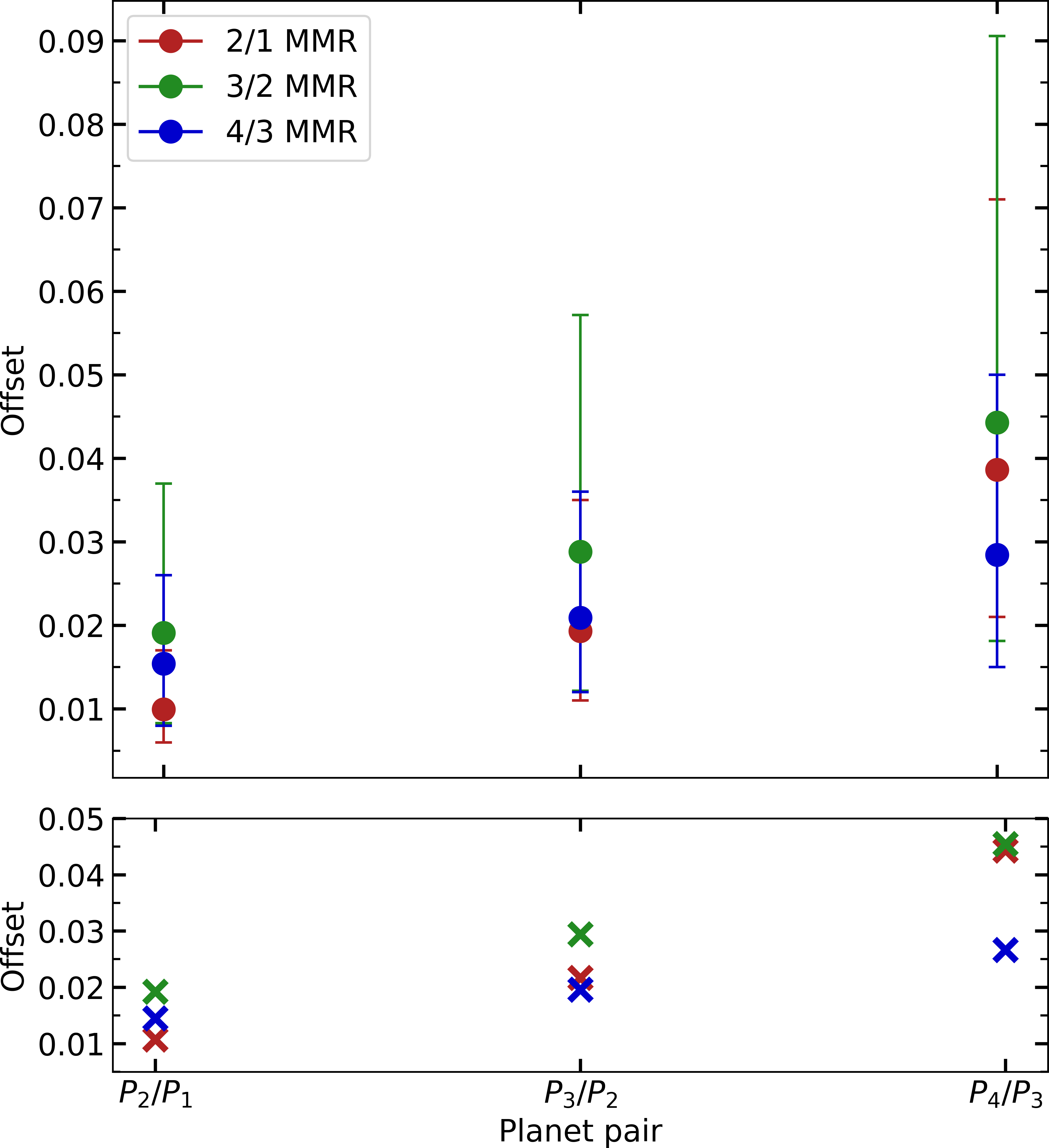}
    \caption{Top panel: Resonance offsets of consecutive pairs for simulations of 4-planet systems captured in 2/1-2/1-2/1 (red), 3/2-3/2-3/2 (green), and 4/3-4/3-4/3 (blue) resonant chains. Bottom panel: For a selected simulation, resonance offsets that the pairs should have if they were in exact 3-planet resonances as expressed in Eq.~\ref{eq:res0}. See text for more details.}
    \label{fig:generic}
\end{figure}
 \begin{figure*}
    \centering
    \includegraphics[width=1.5\columnwidth]{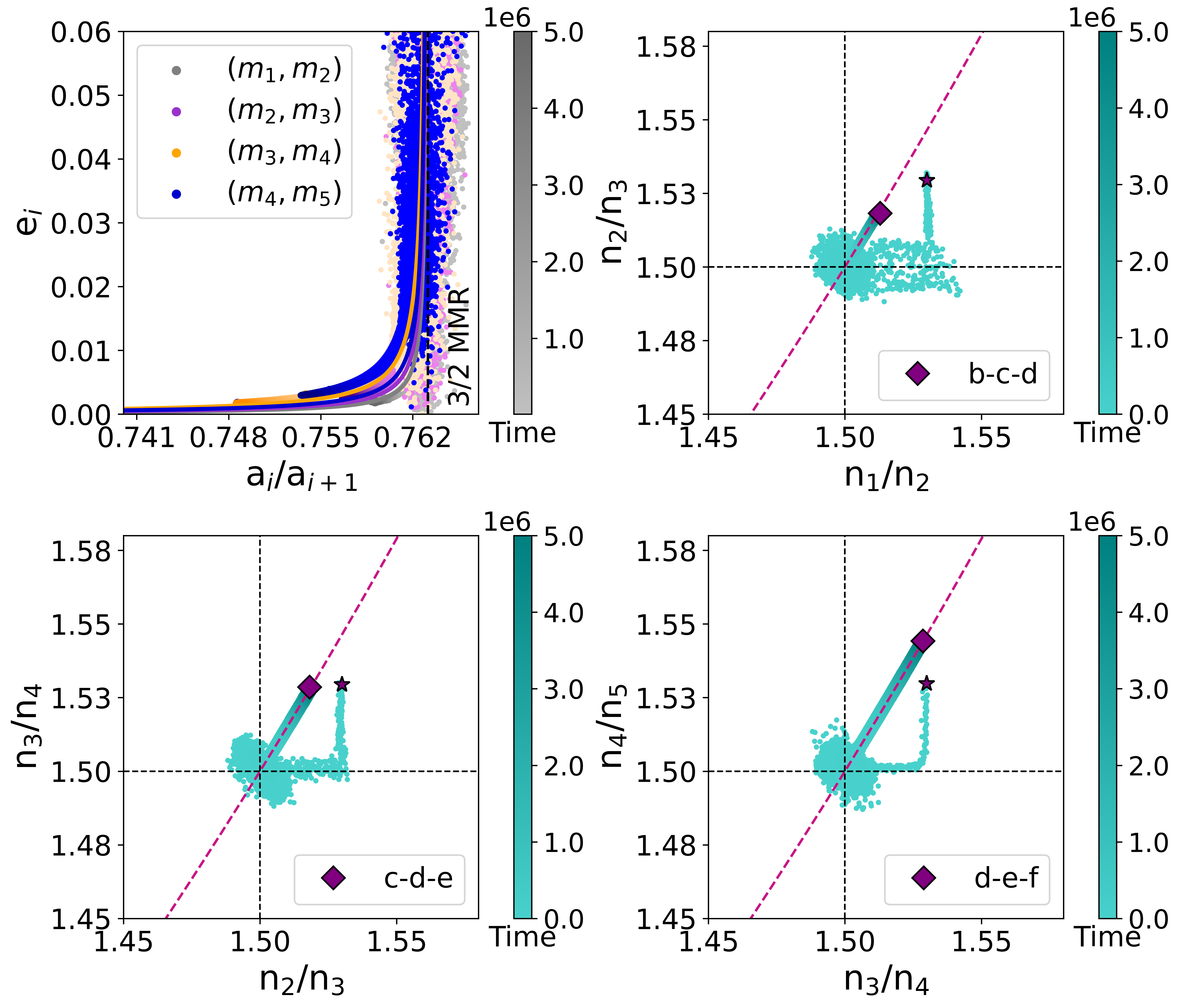}
    \caption{Time evolution of K2-138 as given by the simulation with $Q'=0.1$ on $5 \times 10^6$ yr. Top left panel: evolution of pairs of consecutive planets along the families of periodic orbits for the 3/2 MMR. Other panels: evolution of the planet triplets. See text for more details. }
    \label{fig:peri_K2}
\end{figure*}

In Fig.~\ref{fig:off_sys_sims}, we show the outcomes of the simulations of the late-stage formation and long-term evolution for Kepler-80, Kepler-223, K2-138, TOI-178, and TRAPPIST-1, using the parameters in Table~\ref{tab:sys} with the tidal parameter fixed to $Q'=0.05$ and a fixed integration time of $5 \times 10^6$ yr. The tidal damping timescale deduced from Eq.~\ref{eq:tidal} is displayed in the last column of Table~\ref{tab:sys} for the different systems. As in the previous plots, diamonds represent the resonance offsets for the observed systems, while dots show the resonance offsets given by our simulations. It is clear that the different trends in the offsets described in Section~\ref{sect:obs} are well reproduced by the simulations when considering the tidal effects raised by the star, that in turn follow the chain of resonances that enforces the libration of the Laplace angle predicted from Eq.~\ref{eq:P3_zero}.

\begin{table}
    \centering
    \begin{tabular}{lcc}
    \hline
    System & $Q'$ for $t_{\rm sim}$ & $Q'$ for $t_\star$ \\
    \hline
    Kepler-80  & 0.05 &  4 \\
    Kepler-223 & -    & - \\
    K2-138     & 0.1  &  10 \\
    TOI-178    & 0.01 &  1 \\
    TRAPPIST-1 & 0.05 &  - \\
    \hline
    \end{tabular}
    \caption{$Q'$ values that allow the different resonant chains considered here to reach their observed offsets in a time span of $t_{\rm sim} = 5 \times 10^6$ yr (second column) and a time span equal to $t_\star$ (the age of the system, third column). See text for more details. }
    \label{tab:QforTf}
\end{table}

Regarding the amplitudes of the offsets in Fig.~\ref{fig:off_sys_sims}, they are both in agreement for Kepler-80 and TRAPPIST-1 when using $Q'=0.05$ on $5 \times 10^6$ yr. It is also possible to get the observed offset amplitudes in $5 \times 10^6$ yr for K2-138 and TOI-178 by simply modifying the tidal damping factor as already discussed in  Fig.~\ref{fig:off_m5}. Using $Q'=0.1$, we reproduce both the trend and amplitudes of the offsets for K2-138.
The same holds true for TOI-178 when using the smaller tidal damping value $Q'=0.01$. 
Finally, for Kepler-223 we did not manage to reproduce the offset amplitudes in $5 \times 10^6$ yr, as the planetary eccentricities did not relax towards tidal equilibrium even when considering bigger values of $Q'$. This system is a bit peculiar, since it is much deeper in the resonance than the other systems exhibiting resonant chains. The pairs in Kepler-223 are almost in exact commensurability, i.e., the offsets are almost zero, and possibly the resonant relationship is not entirely governed by the Laplace resonance.
As discussed in \citet{2017A&A...605A..96D}, this might happen due to the order in which the planets get trapped in resonance during migration in our simulations. After that, the tides push the system but not following the 3-planet commensurabilities, and thus it is not possible to reproduce the offset amplitudes. 

Of the five analyzed systems, K2-138 is the only one in which all planet pairs are close to first-order resonance chains following the 3/2-3/2-3/2 MMRs, and the offsets do increase for successive planet pairs. Here we aim to see if the observed increase in the offsets is a general outcome for resonant chains of identical first-order MMRs as it is expected from the analytical formula described in Section \ref{sub:estimation}. To do so, we performed simulations of disk-induced migration and tidal-damped evolution for resonant chains of 4 planets, all in 2/1 MMR or 3/2 MMR or 4/3 MMR. We initially placed the innermost planet at $P_1 = 3.7 \, {\rm days}$ around a $1 \, {\rm M}_\odot$ star and pushed the other 3 planets 3\% outside the resonant commensurability. The planetary masses are randomly chosen within the interval $[2,15] \, {\rm m}_\oplus$. The planets experience a short-term migration ($2 \times 10^4 \, {\rm yr}$) at the same time as they undergo tidal interactions with the star following Eq.~\eqref{eq:tidal} with $Q'=0.05$. After $2 \times 10^4 \, {\rm yr}$ we assume that the disk has dissipated and the planets pursue their long-term evolution dominated by the tides. 

In the top panel of Fig.~\ref{fig:generic}, we show the resonance offsets of 50 systems found in 2/1-2/1-2/1 (red), 3/2-3/2-3/2 (green), and 4/3-4/3-4/3 (blue) resonant chains at the end of the simulation. The dots indicate the mean values after $5 \times 10^6 \, {\rm yr}$, while the $y$-error bars represent the minimal and maximal values found in the 50 simulations of each case. Our simulations confirm that successive pairs captured in the same MMR have increasing resonance offsets as it was expected as a consequence of the preservation of a Laplace resonance. In the bottom panel, we selected one simulation with $P_2/P_1$ near the mean value in the top panel and similarly to Section~\ref{sub:estimation}, we estimated the offsets that the following pairs would have if they were guided by the 3-planet resonance, following Eq.~\ref{eq:P3_zero} for a zero-order commensurability. The agreement between the estimated offsets and the ones found in the simulations is very good, indicating that the 3-planet resonance is the source of the increase in the offsets, independent of the amount of tidal dissipation in the system. The different values of $Q'$ only give different amplitudes of the offsets, but the pattern remains: the offset increases with the distance to the star, when close to a first-order resonant chain. 

Naturally, if the observed offsets are produced by tidal dissipation along 3-planet resonances, the evolution of the system must happen following the corresponding centers of libration. To see the whole picture more clearly, we present in Fig.~\ref{fig:peri_K2} the time evolution of K2-138 where the four planets are all near to 3/2 MMRs. We show the result of a simulation after $5 \times 10^6$ yr with $Q'=0.1$, value for which we find the observed amplitudes of the offsets. In the top left panel, we see the evolution of consecutive pairs of planets along the pericentric branches associated to the 3/2 MMR in the $(a_i/a_{i+1}, e_i)$ representative plane. Each pair evolves on a different branch corresponding to the mass ratio of the planets. We clearly see that the planet pairs follow the family of periodic orbits and thus depart from the nominal resonance. The other three panels in Fig.~\ref{fig:peri_K2} report the time evolution of the system in the $(n_i/n_{i+1},n_{i+1}/n_{i+2})$ planes. The initial conditions of the integration are marked by star symbols and a lighter color that gets darker as time passes until the planets reach their final location. The diamond symbols once again indicate the observed system. The planet triplets first reach a 3/2 MMR between their outer planets, then capture the inner planet in a 3/2 MMR with the middle planet, and finally depart from both MMRs by using the 3-planet resonance as a guide, as previously observed by \citet{2018MNRAS.477.1414C}. 

Throughout the paper we performed the simulations with small values of the tidal quality factor $Q'$ in order to avoid running the simulations for the actual lifetime of the systems. The simulations were stopped after $5 \times 10^6$ yr, moment at which we registered the $Q'$ values that best reproduces the observed offsets for all the systems, and summarized them in the middle column of Table \ref{tab:QforTf}. Now that we have a better understanding of the tidal parameters, we conclude our study computing the dissipation over the age of each system.

More precisely, we conducted simulations similar to those described in the previous section. These simulations involved an initial phase of migration, followed by tidal damping all the way throughout the remaining of the integration. The key distinction is that while the previous simulations had a fixed integration time span of $5 \times 10^6$ yr, the current simulations were adjusted to match the actual age of the planetary system (i.e., $t_\star$ in Table~\ref{tab:sys}). The $Q'$ value that best reproduces the amplitudes in the offsets for the given $t_\star$ integration time span is listed in the last column of Table \ref{tab:QforTf}. 

We find quite similar tidal quality factors for all the systems, since their stellar age are all roughly the same. The consistent $Q'$ values observed across different systems suggest a potential common origin for these offsets. However, it is worth noting that our simulations yield relatively small $Q'$ values compared to the estimated range of $[50, 2500]$ for the Solar System. This highlights our current limited understanding of the dissipation process during MMR evolution.

As previously for $5 \times 10^6$ yr, we were not able to reproduce the offset amplitudes of Kepler-223, and this is why there is no data regarding Kepler-223 in the last column of Table~\ref{tab:QforTf}. A final mention is needed for the case of TRAPPIST-1. Even though we manage to reproduce its offsets with a very simple prescription for tidal dissipation on $5\times 10^6$ yr, we did not succeed to reproduce the system
when running simulations for its whole lifetime. Previous works such as \citet{2022MNRAS.tmp.1824B} and \citet{2022MNRAS.511.3814H}, showed that a single dissipation event cannot reproduce the system as the three inner planets are very far from the original 3/2-3/2 chain while the rest of the planets remain quite close to the commensurabilities in which they were initially captured. When dropping the first planet and performing the simulations with the outer six planets only, we again failed to reproduce the system
on its whole lifetime. Reproducing the architecture of this system is still an open question.

\section{Discussion}
\label{sect:discussion}

In this work, we studied the departure from MMR of five systems with multi-planet resonant chains, namely Kepler-80, Kepler-223, K2-138, TOI-178, and TRAPPIST-1. Using N-body integrations, we simulated their evolution during the migration at the late-stage of the gas phase and after the dispersal of the gas disk where planetary tides raised by the star play an important role. Our goal was not to explore each system individually (and propose a sophisticated formation mechanism for each one), but rather to present for the first time a general study of the preservation of the resonant patterns observed in several systems, with the same general setup.

A quick summary of our findings follows.
\begin{enumerate}
    \item We highlighted the existence of typical trends in the resonance offsets of detected systems with resonant chains.
    \item The observed trends in the offsets are linked with 3-planet resonances and follow analytical estimates given by Eq.~\ref{eq:P3_zero} and Eq.~\ref{eq:P3_one}. As a particular case, the offsets increase for successive planets in resonant chains of identical first-order MMRs, such as the one of K2-138, as a consequence of the preservation of 3-body resonances.
    \item The offset trends of detected systems with resonant chains can be reproduced by tidal damping effects raised by the host star, regardless of the considered value for the tidal factor $Q'$ in Eq.~\ref{eq:tidal}. The planets follow the Laplace resonance when experiencing tidal dissipation and the offsets of the detected systems are consistent with this finding in all test cases.
    \item The amplitudes of the resonant offsets can be reproduced for detected systems with resonant chains when considering appropriate values for the $Q'$ tidal factor and integration timescale. In this respect, $Q'$ determines how long it takes for the system to fully damp and relax towards tidal equilibrium. The offset amplitudes are influenced by the effective tidal dissipation, which can be represented as the total dissipated energy over the integration time, and this scales as $E_{\rm dis} \propto Q' \, t_{\rm sim}$.
    \item For Kepler-80, K2-138, and TOI-178, the amplitudes of the resonant offsets can be reproduced, for the estimated age of the systems, with quite similar tidal quality factor. The consistent $Q'$ values observed across different systems strongly support a common origin for these offsets.
\end{enumerate}

In this work, we showed that the tidal perturbation from the star is a robust mechanism that correctly reproduces the resonance offsets of the planet pairs of Kepler-80, K2-138, TOI-178, and TRAPPIST-1. Kepler-223 would need a more thorough analysis to be accurately reproduced, probably because the system is much deeper in the resonance, leading to tiny offset values.  

Finally, let us mention that further analysis with a more sophisticated model needs to be done. Here we used very simple formulas for both the planetary migration (without inner disk and disk mass decay) and the tidal dissipation (same tidal factor $Q'$ for all the planets) raised by the star. We achieve here the first milestone in the study of offsets for resonant chains, but more realistic prescriptions could be useful to reach a better understanding of their long-term evolution and dynamics, when more systems with resonant chains will be discovered.

\section*{Acknowledgements}
The authors thank Antoine C. Petit for useful comments and discussions that helped us improve our manuscript. This work is supported by the Fonds de la Recherche Scientifique - FNRS under Grant No. F.4523.20 (DYNAMITE MIS-project). Computational resources have been provided by the Consortium des Equipements de Calcul Intensif, supported by the FNRS-FRFC, the Walloon Region, and the University of Namur (Conventions No. 2.5020.11, GEQ U.G006.15, 1610468 et RW/GEQ2016).

\bibliographystyle{aa} 
\bibliography{biblio} 

\end{document}